\documentclass[%
 reprint,
superscriptaddress,
 amsmath,amssymb,
 aps,
]{revtex4-2}

\usepackage{graphicx}
\usepackage{dcolumn}
\usepackage{bm}

\usepackage[dvipsnames]{xcolor}

\begin{document}

\preprint{APS/123-QED}

\title{A Generalized Einstein Relation for\\ Markovian Friction Coefficients from Molecular Trajectories}

\author{J. M. Hall}
\email{jhall4@uoregon.edu, jmonroe.hall4@gmail.com}

 \affiliation{%
 Department of Physics, University of Oregon, Eugene, Oregon 97403, USA}%

\author{M. G. Guenza}%
 \email{mguenza@uoregon.edu}

\affiliation{%
 Department of Chemistry and Biochemistry, University of Oregon, Eugene, Oregon 97403, USA
}%

\date{\today}

\begin{abstract}
We present a generalized Einstein relation for the friction coefficients associated with an underlying memory kernel in terms of observable time correlation functions. There is considerable freedom in the correlations involved, and this allows the expression to be tailored to the particular system to achieve numerical stability. We demonstrate this by recovering the site-specific friction coefficients from trajectories of a freely diffusing model trimer, and we show that the accuracy is greatly improved over established Volterra inversion methods for kernel extraction.
\end{abstract}

\maketitle


Coarse-grained (CG) dynamical models of complex systems aim to explicitly describe a reduced set of quantities of interest while only implicitly modeling the remaining degrees of freedom. For example, a protein in solution may be described by the centers-of-mass of the constituent amino acid monomers, and its dynamics described by a Langevin equation, where the solvent and atomistic degrees of freedom are modeled as friction and noise. Such CG models have the potential to provide analytical insight into the mechanisms behind biological function of macromolecules on targeted length scales \cite{Dill2012,Copperman2017,Socci1996b,Lange2008a} as well as to increase computationally accessible timescales by projecting out high-frequency modes.

The Mori-Zwanzig projection operator method can be used to rigorously derive a Generalized Langevin Equation (GLE) for the time evolution of the CG variables consistent with an underlying atomistic Hamiltonian \cite{Mori1965,Zwanzig2001}. Recent years have seen a growing interest in developing GLEs as a practical tool due to the modern accessibility of long trajectories via atomistic Molecular Dynamics (MD) simulations and new methods of parameterizing the equations of motion using statistical correlations from these trajectories \cite{kowalikNetzExtraction, daltonFoldingGovernedMemory, leeDarveMultidimAlanine}.

Despite these advances and the growing interest in designing fast CG computational methods to simulate complex systems across multiple scales, parametrizing the GLE with high accuracy remains an ongoing challenge. A key task in designing a highly accurate GLE is the calculation of the friction coefficients entering the CG model.\cite{Straub,Berne,Daldrop,Daldrop1}  In this manuscript, we present a solution for the friction coefficients in the Markovian regime by proposing an extended Einstein relation consistent with an underlying memory function. We evaluate the quality of our approach on a simple three-bead model, and compare the accuracy of the recovered friction with that of a conventional method involving the numerical inversion of a Volterra integral equation.

We adopt this simplified model with a known analytical, multidimensional free energy landscape, because the accuracy of the friction coefficient depends in our model on the accuracy of the conservative forces. Extensive literature on deriving diffusion coefficients (friction)  from simulation trajectories\cite{Berkowitz} or experimental data\cite{de Oliveira, Foster,Freitas,Lannon,Satija} of more complex systems, such as protein folding, commonly identifies a single or a pair of reaction (collective) coordinates.\cite{Hummer,Best,Best1} Rare events are typically analyzed along these coordinates using enhanced sampling of the free energy landscape.\cite{Mugnai} Testing our method on such complex systems would introduce uncertainties on the intramolecular forces, which would cloud the quality of our results.\cite{Best,Best1,Hinczewski} Therefore, we rely on a simple three-bead model with known interactions. This model has a multidimensional free energy landscape, both bound and unbound coordinates, providing a simplified representation of the complexity inherent to the dynamics of a coarse-grained macromolecule.

The proposed model for calculating self and cross friction coefficients provides a significant advantage over traditional methods by allowing flexibility in the selection of vectors entering the Generalized Einstein Relation. This adaptability enables one to resolve instabilities in the friction coefficients by selecting appropriate coordinates.

In the Zwanzig-projected equation there are two functions that must be parameterized. One is the potential of mean force $U$, which can be obtained from equilibrium statistics. The other is the memory function $\Gamma$, which in general is a function of time and the CG positions and momenta \cite{hijonZwanzig}. In this work we take the common approximation that this kernel is state-independent, leaving only the time-dependent memory.

If the CG coordinates are taken to be the center-of-mass positions of $n$ distinct clusters of atoms, then the state-independent form of the Zwanzig GLE for cluster $i$ is given by
\begin{eqnarray}
\dot{\vec{r}}_{i}(t)=&&\frac{1}{m_{i}}\vec{p}_{i}(t) \nonumber
\\
\dot{\vec{p}}_{i}(t)=&& -\frac{\partial U(\vec{r}(t))}{\partial \vec{r}_{i}}-\sum_{j}\int_{0}^{t}d\tau \: \Gamma_{ij}(t-\tau)\vec{v}_{j}(\tau) +\vec{f}^{R}_{i}(t), \nonumber
\\
\label{eq:GLE}
\end{eqnarray}
where $\vec{r}_{i}$ is the position of the $i$-th center of mass, $\vec{p}_i$ and $\vec{v}_{i}$ are the corresponding momentum and velocity, and $\vec{f}^{R}$ is a set of gaussian-distributed random forces whose statistics are related to $\Gamma$ by the second fluctuation-dissipation theorem \cite{VanKampen2007},
\begin{equation}
\langle f^{R}_{i\alpha}(t) f^{R}_{j\beta}(t')\rangle = k_{B}T\Gamma_{ij}(t-t')\delta_{\alpha\beta}, \label{fluctuationDissipation}
\end{equation}
where $T$ is the temperature and $k_{B}$ is the Boltzmann constant. The Greek indices indicate Cartesian components and the Kronecker delta reflects our assumption of an isotropic environment. 

When the memory function is integrable, as is typical, this time-dependent kernel can be replaced by a constant friction coefficient to model kinetics over sufficiently long timescales, where the behavior is Markovian. This Markovian regime is of particular interest for its mathematical simplicity compared to the full GLE when accurate dynamics over the shortest timescales are not crucial. This is often the case in the study of diffusive motion in macromolecules, where relaxation of the solvent and other local degrees of freedom can occur much faster than conformational changes of functional interest. 

In the Markovian limit, the running integral $G(t)$ of the memory function should converge to a stable plateau yielding the friction coefficients
\begin{equation}
\zeta =  \lim_{t\to\infty} G(t)  = \lim_{t\to\infty} \int_{0}^{t}d\tau\:\Gamma(\tau).
\label{friction}
\end{equation}

The stability and accuracy of this plateau at long lag times is then crucial to developing Markovian CG models from trajectory data. This is especially so in the case of multidimensional models that possess a rich mode structure over a wide hierarchy of timescales. Correlations in the fastest coordinates may completely decay while others are still in a non-Markovian regime, and such coordinates may be coupled via off-diagonal friction through mechanisms such as hydrodynamic interactions. Such cases require an approach that is numerically reliable at long timescales despite the exponential decay of the input time correlation functions (TCFs).

One popular method to directly extract the memory kernel from MD trajectories involves the inversion of a Volterra integral equation of the first kind \cite{Berne,Berkowitz}. The friction coefficients can then be obtained from the total integral of this memory function. We now show that this method can lead to instability in the friction coefficients and propose a new method that yields a far more accurate and stable plateau at long timescales.

By taking the dot product of the force equation in (\ref{eq:GLE}) with the initial velocity $\vec{v}_{k}(0)$ and then taking the mean, one obtains for a multi-particle model in an isotropic solvent \cite{kowalikNetzExtraction}
\begin{equation}
M\dot{C}^{vv}(t)= -C^{vU'}(t)-\int_{0}^{t}dt \: \Gamma(t-\tau)C^{vv}(\tau), \label{eq:vel_avg_GLE}
\end{equation}
where 
\begin{equation}
C^{xy}_{ij}(t) := \langle\vec{x}_{i}(0)\cdot\vec{y}_{j}(t)\rangle
\label{eq:correlation_def}
\end{equation}
is the time-lagged correlation function between $\vec{x_{i}}$ and $\vec{y}_{i}$, and $M$ is the diagonal matrix of cluster masses. As a projected CG coordinate, $\vec{v}_{k}$ is uncorrelated with the random noise $\vec{f}^{R}_i$ for all time lags (see Supplemental Material (SM) \cite{supplementalMaterial}). Integrating Eq. (\ref{eq:vel_avg_GLE}) over time yields
\begin{equation}
MC^{vv}(t) - MC^{vv}(0) = \mathcal{F}^{vU'}(t)-\int_{0}^{t}d\tau \: G(t-\tau)C^{vv}(\tau), \label{eq:integrated_GLE}
\end{equation}
where $\mathcal{F}^{vU'}(t)$ is defined as
\begin{equation}
\mathcal{F}^{vU'}(t) =-\int_{0}^{t}d\tau\:C^{vU'}(\tau) \label{eq:F_def}
\end{equation}
and $G(t)$ is the integral of the memory function, Eq.(\ref{friction}).

Conventionally, the integral in Eq. (\ref{eq:integrated_GLE}) is discretized in order to solve for $G(t)$. The trapezoid rule can be used for this discretization, but leads to spurious high-frequency oscillations in the solution which must be filtered out \cite{linzNumericalVolterra}. We use, instead, the left-handed rectangle rule, which after some rearranging transforms Eq. (\ref{eq:integrated_GLE}) into
\begin{align}
G(t) =& \frac{1}{\Delta \tau}\left[ \mathcal{F}^{vU'}(t) + MC^{vv}(0) - MC^{vv}(t) \right]C^{vv^{-1}}(0) \notag \\ &+ \sum_{i=1}^{N-1}G(t-i\Delta \tau)C^{vv}(i\Delta \tau)C^{vv^{-1}}(0)
\label{eq:discretized_GLE}
\end{align}
where $N=t/\Delta\tau$. Once the correlation functions on the right-hand side are measured, Eq. (\ref{eq:discretized_GLE}) can be solved iteratively for $G(t)$ starting from $G(0) = 0$. 

Variations of this procedure have been used frequently in the literature to calculate $G(t)$ for many single-dimensional CG models of molecular systems, and occasionally for multi-dimensional models. If a Markovian description is desired, then the limit of $G(t)$ as $t\rightarrow\infty$ furnishes the Markovian friction coefficients. If a non-Markovian description is necessary, then $G(t)$ can be differentiated to obtain the memory function, $\Gamma(t)$.

Unfortunately, since TCFs typically decay exponentially and are calculated from finite MD simulations, a low ratio of signal to noise becomes inevitable at long lag times, leading to poor stability of the long-time plateau in the integral. This limits the applicability of the Volterra inversion method to calculating Markovian friction coefficients.

To illustrate the memory extraction problem and the limitations of the procedure described by Eq. (\ref{eq:discretized_GLE}), we simulate Eq. (\ref{eq:GLE}) for a toy molecule. The molecule is formed by three CG sites, interacting with a known internal potential and memory function. The accuracy of the procedure is tested by attempting to recover the known friction coefficients via the integral of the memory function.

The internal potential of the trimer is given by
\begin{equation}
U(l_{1},l_{2},\theta) = \frac{k_{1}}{2}(l_{1}-l_{1_{0}})^{2} + \frac{k_{2}}{2}(l_{2}-l_{2_{0}})^{2} + \frac{k_{\theta}}{2}(\theta-\theta_{0})^{2} \label{eq:toy_potential}
\end{equation}
where $l_{1}, l_{2}$ are the bond lengths along the three-bead chain and $\theta$ is the angle between them, and $(l_{1_{0}}, l_{2_{0}}, \theta_{0})$ is the minimum-energy configuration, which we set equal to $(1,1,\frac{\pi}{2})$. The spring constants are set to $(k_{1}, k_{2}, k_{\theta}) = (14, 20, 7)$.

The memory function of the trimer is taken to be
\begin{equation}
\Gamma_{ij}(t) = 10e^{-10t/\zeta_{ij}} \label{eq:toy_memory}
\end{equation}
and the Markovian friction coefficients are
\begin{equation}
\zeta =\begin{bmatrix}
10 & 0 & 10\\
0 & 10 & 0\\
10 & 0 & 20
\end{bmatrix}
\end{equation}
indicating a single off-diagonal interaction between the ends of the trimer. The beads in the trimer are assigned masses $m = (30,40,30)$.

The time-correlated noise is generated according to Eq. (\ref{fluctuationDissipation}) using the method detailed in \cite{garciaOljavoSancho_noise}. As the sum of many interactions, the noise is taken to be gaussian-distributed, motivated by the Central Limit Theorem. Equation (\ref{eq:GLE}) is then simulated using the Euler method with time step $\Delta t = 0.01$ for a total time of $T_{sim} = 10^{4}$.

\begin{figure}[b]
\includegraphics[width = 8.6cm]{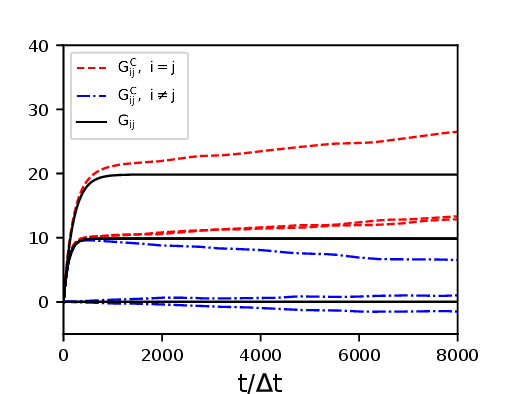}
\caption{\label{fig:memoryExtract} The diagonal (dashed) and off-diagonal (dot-dashed) components of the calculated $G^{C}(t)$ as obtained via the Volterra inversion method compared to the exact solution $G(t)$ obtained from Eq. \ref{eq:toy_memory} (solid).}
\end{figure}

In Fig. \ref{fig:memoryExtract}, we compare the results of extracting the memory function from the resulting trajectory using equation (\ref{eq:discretized_GLE}) with the true integral of the known $\Gamma$, Eq. \ref{eq:toy_memory}. While the Volterra inversion method matches $G(t)$ well for short times, it begins to drift at timescales comparable to the Markov timescale we are interested in, and the error becomes progressively worse at longer timescales. Without a stable plateau, there is no unambiguous result for the friction coefficients.

To address these shortcomings of the Volterra inversion method, we propose an alternative approach that circumvents the direct calculation of $\Gamma(t)$ while reliably yielding the plateau value.

First we rewrite Eq. (\ref{eq:integrated_GLE}) as
\small
\begin{equation}
MC^{vv}(t) - MC^{vv}(0) = \mathcal{F}^{vU'}(t)-\int_{0}^{t}dt'\int_{0}^{t'}d\tau \: \Gamma(t'-\tau)C^{vv}(\tau).
\label{eq:integrated_GLE_2}
\end{equation}
\normalsize

We then define
\begin{equation}
\mathcal{D}^{vv}(t) := \int_{0}^{t}d\tau \:C^{vv}(\tau) \ , \label{eq:D_def}
\end{equation}
and multiply both sides of equation (\ref{eq:integrated_GLE_2}) by $\mathcal{D}^{vv^{-1}}$ to get

\begin{equation}
\zeta_{v}(t) = \left[ MC^{vv}(0) - MC^{vv}(t) + \mathcal{F}^{vU'}(t) \right] {\mathcal{D}^{vv^{-1}}}(t)\label{eq:zeta_v_formula}
\end{equation}
where we have defined

\begin{equation}
\zeta_{v}(t) := \left[\int_{0}^{t}dt'\int_{0}^{t'}d\tau \: \Gamma(t'-\tau)C^{vv}(\tau) \right] {\mathcal{D}^{vv^{-1}}}(t)\label{eq:zeta_v_def} \ .
\end{equation}

The subscript $v$ recalls the initial value we multiplied across Eq. (\ref{eq:GLE}). Since the inner $\tau$ integral is a convolution between $\Gamma$ and $C^{vv}$, in the limit $t\rightarrow\infty$ the double integral separates into two factors 

\begin{equation}
\zeta_{v}(\infty) = \left[\int_{0}^{\infty}dt \: \Gamma(t) \int_{0}^{\infty}dt \: C^{vv}(t) \right] {\mathcal{D}^{vv^{-1}}}(\infty)\label{eq:zeta_v_separation}
\end{equation}

\noindent which by the definitions of $G$ and $\mathcal{D}$ simplifies to

\begin{equation}
\zeta_{v}(\infty) = G(\infty)\label{eq:zeta_v_limit} \ .
\end{equation}

We have in $\zeta_{v}(t)$ a quantity distinct from the integral of the memory function that nonetheless has the same limit at long times. Using Eq. (\ref{eq:zeta_v_formula}) the Markovian friction coefficients can be calculated directly from observed TCFs without solving an integral equation while maintaining consistency with the underlying memory function. To identify the Markovian timescale $T_{M}$, we calculate $\zeta_{v}(t)$ for multiple values of $t$ in order to monitor convergence.

In the long time limit, one can understand Eq. (\ref{eq:zeta_v_formula}) as a generalized Einstein relation. If we consider the case of a single molecular center-of-mass diffusing freely such that $U'(\vec{r})=0$, we have

\begin{equation}
\zeta_{v}(\infty) = \frac{mC^{vv}(0)}{ {\mathcal{D}^{vv}}(\infty)}\label{eq:einstein_1}
\end{equation}

\noindent given that $\mathcal{F}^{vU'}(t) = 0$ for all $t$, and $C^{vv}(t)\rightarrow 0$ as $t\rightarrow \infty$. By the equipartition theorem, one has $mC^{vv}(0) = 3k_{B}T$. The usual diffusion coefficient $D$ is identified via $\mathcal{D}^{vv}(\infty)=3D$, yielding the Einstein relation

\begin{equation}
\zeta_{v}(\infty) = \frac{k_{B}T}{D}\label{eq:einstein_2} \ .
\end{equation}

The proposed formalism, illustrated in Eq. (\ref{eq:zeta_v_formula}), is a convenient method to calculate the friction coefficients in systems described entirely by unbound coordinates, such as the center-of-mass diffusion of a free molecule. When attempting to model the internal dynamics of a molecule, however, a further obstacle arises that requires some generalization of the approach.

For a single bound coordinate, such as an internal coordinate of a coarse-grained macromolecule, the mean-squared displacement saturates to a finite value over time instead of growing linearly. In this case, the diffusion coefficient $\lim_{t\rightarrow\infty}\mathcal{D}^{vv}(t) = 0$. When modeling $n$ atom clusters to describe an entire molecule, only the eigenvalue associated with translation of the entire CG molecule has a finite limit, while the remaining $n-1$ eigenvalues go to zero at long times. 

The numerator of $\zeta_{v}(t)$ goes to a finite value at large $t$ (see SM \cite{supplementalMaterial}), and $\zeta_{v}(\infty)$ diverges. The divergence of the Markovian limit in the case of bound coordinates was noted and resolved in Ref. \cite{leiParamUpdate} for a different approach to the memory function using a hierarchy of approximations. Here we implement a solution inspired by that approach that also simultaneously handles the unbound translational mode.

The formalism we derived in Eq. (\ref{eq:zeta_v_formula}), can be easily generalized to solve this issue. While we initially chose to take the dot product of Eq. (\ref{eq:GLE}) with $\vec{v}_{k}(0)$, we now generalize this approach by taking the dot product with an arbitrary vector-valued function of the initial CG phase space variables $\vec{g}_{k}(\vec{r}(0),\vec{p}(0))$. To the extent that the state-independent GLE is valid, this vector $\vec{g}_{k}$ remains orthogonal to the noise since all CG variables are Zwanzig-projected coordinates (see SM \cite{supplementalMaterial}). The resulting generalized integral equation is:

\begin{equation}
M\dot{C}^{gv}(t)= -C^{gU'}(t)-\int_{0}^{t}d\tau \: \Gamma(t-\tau)C^{gv}(\tau). \label{eq:avg_GLE_g}
\end{equation}

Integrating in the same manner as before,

\begin{equation}
\zeta_{g}(t) = \left[ MC^{gv}(0) - MC^{gv}(t) + \mathcal{F}^{gU'}(t) \right] {\mathcal{D}^{gv^{-1}}}(t),\label{eq:zeta_g_formula}
\end{equation}

\noindent yields a generalized friction coefficient

\begin{equation}
\zeta_{g}(t) := \left[\int_{0}^{t}dt'\int_{0}^{t'}d\tau \: \Gamma(t'-\tau)C^{gv}(\tau) \right] {\mathcal{D}^{gv^{-1}}}(t).\label{eq:zeta_g_def}
\end{equation}

We note that
\begin{equation}
\zeta_{g}(\infty) = G(\infty) = \zeta\label{eq:zeta_g_limit}
\end{equation}
for all choices of $\vec{g}$. In the long time limit, Eq. (\ref{eq:zeta_g_formula}) then provides an even more generalized form of the Einstein relation, connecting the friction coefficient with ratios of a broader class of correlations resulting from different choices of $\vec{g}$.

We are now free to search for any set of vectors $\vec{g}_{k}(\vec{r},\vec{p})$ such that ${\mathcal{D}^{gv}}$ retains a full set of finite eigenvalues in the limit $T\rightarrow\infty$. For the case where all positions are bound, \cite{leiParamUpdate} were able to obtain a stable Markov limit using correlations equivalent to setting $\vec{g}_{k} = \vec{r}_{k}$ here. Inspired by their work we adopt

\begin{equation}
    \vec{g}_{0}:=\\
    \begin{bmatrix} 
   \vec{r'}_{C}/ \tau_0\\
    \\
    \vec{V}\\
    \end{bmatrix}\label{eq:g_trial}
\end{equation}

\noindent where $\vec{V}$ is the center-of-mass velocity and $\vec{r'}_{C}/\tau_0$ is the vector of CG position coordinates relative to the center-of-mass, normalized by the unit of time, $\tau_0$. Here, the $x,y,z$ coordinates of the $N$th CG site are removed, such that $\vec{r'}_{C} = (x^{C}_1,y^{C}_1,z^{C}_1, ... , x^{C}_{n-1}, y^{C}_{n-1}, z^{C}_{n-1})$. The motivation is to associate the bound modes of rotational and internal vibration with the bound position coordinates, $\vec{r}_{C}$, relative to the center-of-mass position, where one set of coordinates has been removed to maintain the correct vector's size. However, it is essential to include the unbound translational mode of the center of mass. If we simply set $\vec{g} = \vec{r}_{C}$, the bound position coordinates exclude the translational degrees of freedom, leading to an unwanted zero eigenvalue of $\mathcal{D}^{gv}(t)$. 

Finally, we observe that while $\mathcal{D}^{gv}(t)$ has finite eigenvalues as $t\rightarrow\infty$ for $\vec{g}=\vec{g}_{0}$, its derivative $C^{gv}(t)$ now has zero eigenvalues at $t=0$, as dictated by the equipartition theorem. Given that the first derivative of the numerator in equation (\ref{eq:zeta_g_formula}) is finite, L'Hôpital's rule indicates that the limit as $t\rightarrow 0$ becomes infinite (see SM \cite{supplementalMaterial}). By adopting  $\vec{g}=\vec{g}_{0}$ we effectively exchanged the long-time instability for a short-time one. Although this does not strictly pose a problem for the calculation of the Markovian friction, it would be preferable to achieve stability at both limits. This issue can be easily resolved through an additional adjustment of our $\vec{g}$-vector,

\begin{equation}
    \vec{g}:=\\
    \begin{bmatrix} 
   \vec{r'}_{C}/ \tau_0 - \vec{v'}_{L} \\
    \\
    \vec{V}\\
    \end{bmatrix}\label{eq:g_final}
\end{equation}

\noindent where $\vec{v'}_{L}$ are the lab frame velocities where one set of velocity coordinates is removed. Since $C^{vv}(0)$ possesses a full set of finite eigenvalues due to the equipartition theorem, this stabilizes the $t\rightarrow 0$ limit (see SM \cite{supplementalMaterial}). Hence, Eq. (\ref{eq:g_final}) will be our choice for $\vec{g}$ from this point on.

\begin{figure}[b]
\includegraphics[width = 8.6cm]{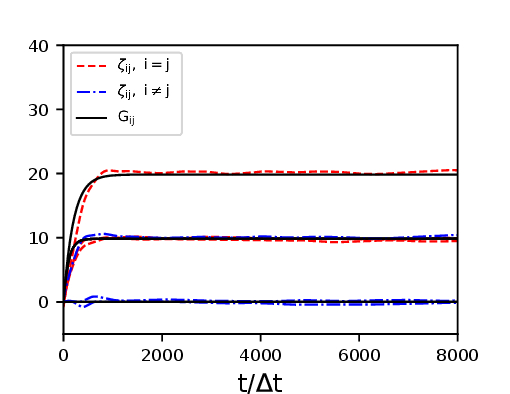}
\caption{\label{fig:zetaExtract} The diagonal (dashed) and off-diagonal (dot-dashed) components of $\zeta_{g}(T)$ compared with the exact components of $G(T)$ (solid).}
\end{figure}

We now calculate $\zeta_{g}(t)$ via Eqs. (\ref{eq:zeta_g_formula}, \ref{eq:g_final}) using the same trajectory of the trimer as before. In principle the Markovian friction can be obtained by calculating $\zeta_{g}(t)$ at a single time $t>T_{M}$. In practice, however, $T_{M}$ is generally unknown. In this case $\zeta_{g}(t)$ can be calculated at multiple sparse values of $t$ and monitored for convergence. In Fig. \ref{fig:zetaExtract} we plot $\zeta_{g}(t)$ at full resolution to facilitate comparison with $G(t)$.

The results demonstrate a clearly identifiable and stable plateau that agrees well with the long-time limit of the exact $G(t)$.

We have presented an Einstein relation generalized to systems of interacting particles, where the friction coefficients are consistent with the total integral of an underlying memory function. Due to the nature of the Einstein relation as a ratio, it is possible to further generalize to a broader class of correlations than the usual velocity-velocity correlations corresponding to the diffusion coefficient, and all of these ratios yield the same friction coefficients. The generality of this formalism allows us the freedom of selecting appropriate coordinates to resolve instabilities in the memory integral.

This relation could serve as a powerful tool for constructing Markovian CG models. Not only does it yield accurate friction coefficients without explicit calculation of the memory function, but its flexibility allows it to be tailored to the coordinate structure of the system at hand in order to achieve robust numerical stability. The example we have provided demonstrates how site-specific friction coefficients simultaneously consistent with bound internal dynamics and unbound global diffusion in a free molecule can be obtained.

\vspace{4mm}

\textit{Acknowledgements-} This material is based upon work supported by the National Science Foundation under Grant No. CHE-2154999. The work was also supported by the Molecular Biology and Biophysics Training Program (MBBTP) under the Institute of Molecular Biology at the University of Oregon.
The computational work was partially performed on the
supercomputer Expanse at the San Diego Supercomputer Center, with the support of ACCESS \cite{access} allocation Discover ACCESS CHE100082 (ACCESS is a program supported by the National Science Foundation under Grant No. ACI-1548562). This work also benefited from access to the University of Oregon high performance computing cluster, Talapas. We thank Alex Batelaan for the careful reading of the manuscript.


\providecommand{\noopsort}[1]{}\providecommand{\singleletter}[1]{#1}%


\begin{thebibliography}{18}%
\makeatletter
\providecommand \@ifxundefined [1]{%
 \@ifx{#1\undefined}
}%
\providecommand \@ifnum [1]{%
 \ifnum #1\expandafter \@firstoftwo
 \else \expandafter \@secondoftwo
 \fi
}%
\providecommand \@ifx [1]{%
 \ifx #1\expandafter \@firstoftwo
 \else \expandafter \@secondoftwo
 \fi
}%
\providecommand \natexlab [1]{#1}%
\providecommand \enquote  [1]{``#1''}%
\providecommand \bibnamefont  [1]{#1}%
\providecommand \bibfnamefont [1]{#1}%
\providecommand \citenamefont [1]{#1}%
\providecommand \@href[1]{\@@startlink{#1}\@@href}%
\providecommand \@@href[1]{\endgroup#1\@@endlink}%
\providecommand \@sanitize@url [0]{\catcode `\\12\catcode `\$12\catcode `\&12\catcode `\#12\catcode `\^12\catcode `\_12\catcode `\%12\relax}%
\providecommand \@@startlink[1]{}%
\providecommand \@@endlink[0]{}%
\providecommand \urlprefix  [0]{URL }%
\providecommand \doibase [0]{https://doi.org/}%
\providecommand \selectlanguage [0]{\@gobble}%
\providecommand \bibinfo  [0]{\@secondoftwo}%
\providecommand \bibfield  [0]{\@secondoftwo}%
\providecommand \translation [1]{[#1]}%
\providecommand \BibitemOpen [0]{}%
\providecommand \bibitemStop [0]{}%
\providecommand \bibitemNoStop [0]{.\EOS\space}%
\providecommand \EOS [0]{\spacefactor3000\relax}%
\providecommand \BibitemShut  [1]{\csname bibitem#1\endcsname}%
\let\auto@bib@innerbib\@empty
\bibitem{Dill2012}{K. A. Dill and J. L. MacCallum, Science \textbf{338}, 1042 (2012).}

\bibitem{Copperman2017}{J. Copperman, M. Dinpajooh, E. R. Beyerle, and M. G. Guenza, Phys. Rev. Lett. \textbf{119}, 158101 (2017).}

\bibitem{Socci1996b}{N. D. Socci, J. N. Onuchic, and P. G. Wolynes, J. Chem. Phys. \textbf{104}, 5860 (1996).}

\bibitem{Lange2008a}{O. F. Lange and H. Grubmuller, Proteins: Structure, Function and Genetics \textbf{70}, 1294 (2008).}

\bibitem{Mori1965}{H. Mori, Prog. Theor. Phys., Japan \textbf{33}, 423 (1965).}

\bibitem{Zwanzig2001}{R. Zwanzig, \textit{Nonequilibrium statistical mechanics} (Oxford : New York : Oxford University Press, 2001).}

\bibitem{kowalikNetzExtraction}{B. Kowalik, J. O. Daldrop, J. Kappler, J. C. F. Schulz, A. Schlaich, and R. R. Netz, Phys. Rev. E \textbf{100}, 012126
(2019).}

\bibitem{daltonFoldingGovernedMemory}{B. A. Dalton, C. Ayaz, H. Kiefer, and R. R. Netz, Proc. Natl. Acad. Sci. \textbf{120}, 31 (2023).}

\bibitem{leeDarveMultidimAlanine}{H. S. Lee, S. Ahn, and E. F. Darve, J. Chem. Phys. \textbf{150},
174113 (2019).}

\bibitem{Straub}{J. E. Straub, M. Borkovec, and B. J. Berne, J. Phys. Chem. \textbf{91}, 4995 (1987).}

\bibitem{Berne}{B. J. Berne and G. D. Harp, in \textit{Advances in Chemical Physics} (John Wiley \& Sons, Ltd, 1970) Chap. 3, p. 63.} 

\bibitem{Daldrop}{J. O. Daldrop, J. Kappler, F. N. Brunig, and R. R. Netz,
Proc. Natl. Acad. Sci. U. S. A. \textbf{115}, 5169 (2018).}

\bibitem{Daldrop1}{J. O. Daldrop, B. G. Kowalik, and R. R. Netz, Phys. Rev. X \textbf{7},
041065 (2017).}


\bibitem{Berkowitz}{M. Berkowitz, J. D. Morgan, D. J. Kouri, and J. A. McCammon, J. Chem. Phys. \textbf{75}, 2462 (1981).}

\bibitem{Foster}{D. A. N. Foster, R. Petrosyan, A. G. T. Pyo, A. Hoffmann, F. Wang, and 
M. T.  Woodside, Biophys. J.
\textbf{114}, 1657 (2018).}

\bibitem{de Oliveira}{R. J. de Oliveira, J.
Chem. Phys. \textbf{149}, 234107 (2018).}

\bibitem{Freitas}{F. C. Freitas, A. N. Lima, V. G. Contessoto, P. C. Whitford, and R. J.
Oliveira, J. Chem. Phys. \textbf{151}, 114106 (2019).}

\bibitem{Lannon}{H. 
Lannon, J. S. Haghpanah, J. K. Montclare, E. Vanden- Eijnden, and J.
Brujic, Phys. Rev.
Lett. \textbf{110}, 128301 (2013).}

\bibitem{Satija}{R. Satija, and D. E. Makarov, J. Phys. Chem. B \textbf{123}, 802 (2019).}

\bibitem{Hummer} {G. Hummer, New J. Phys. \textbf{7}, 34 (2005).}

\bibitem{Best}{
R. B. Best, and G. Hummer, Proc. Natl. Acad. Sci. U. S. A. \textbf{107}, 1088 (2010).}

\bibitem{Best1}{R. B. Best, and G. Hummer, Phys.
Chem. Chem. Phys. \textbf{13}, 16902 (2011).}

\bibitem{Mugnai}{M. L. Mugnai, and R. Elber, J. Chem. Phys.
\textbf{142}, 014105 (2015).}

\bibitem{Hinczewski}{
M. Hinczewski, Y. von Hansen, J. Dzubiella, and R. R. Netz,  J. Chem. Phys. \textbf{132}, 245103 (2010).}

\bibitem{hijonZwanzig}{C. Hijon, P. Espanol, E. Vanden-Eijnden, and R. Delgado-Buscalioni, Faraday Discuss. \textbf{144}, 301 (2010).}

\bibitem{VanKampen2007}{N. G. Van Kampen, \textit{Stochastic Processes in Physics and Chemistry, Third Edition} (Elsevier, 2007).}

\bibitem{supplementalMaterial}{See Supplemental Material [url] for proof of the orthogonality between functions of the CG variables and the noise, proof that the numerator of $\zeta_v$ goes to a finite limit, and plots of the resulting friction calculations using alternate choices of $\vec{g}$ mentioned but not illustrated in the main text.}

\bibitem{linzNumericalVolterra}{P. Linz, The Computer Journal \textbf{12}, 393 (1969).}

\bibitem{garciaOljavoSancho_noise}{J. Garcıa-Ojalvo and J. M. Sancho, \textit{Noise in Spatially Extended Systems} (Springer-Verlag New York, Inc., 1999)
pp. 95–97.}


\bibitem{leiParamUpdate}{F. Grogan, H. Lei, X. Li, and N. A. Baker, J. Chem.Phys. \textbf{418}, 109633 (2020).}

\bibitem{access}{T. J. Boerner, S. Deems, T. R. Furlani, S. L. Knuth, and J. Towns, ACCESS: Advancing Innovation: NSF’s Advanced Cyberinfrastructure Coordination Ecosystem: Services \& Support, in \textit{Practice and Experience in Advanced Research Computing}, PEARC’23 (Association for Computing Machinery, New York, NY, USA, 2023) p. 173–176.}







\end{thebibliography}
\end{document}